\documentclass[preprintnumbers,superscriptaddress,amsmath,amssymb,nofootinbib,twocolumn,showpacs]{revtex4-1}
\usepackage{mathtools}
\usepackage{graphicx}
\usepackage{color}
\usepackage{natbib}
\usepackage{multirow}
\usepackage{SIunits}
\usepackage{xcolor} 

\newcommand{\beq}{\begin{equation}}
\newcommand{\eeq}{\end{equation}}
\newcommand{\ben}{\begin{eqnarray}}
\newcommand{\een}{\end{eqnarray}}
\newcommand{\besub}{\begin{subequations}}
\newcommand{\eesub}{\end{subequations}}
\newcommand{\bi}{\begin{itemize}}
\newcommand{\ei}{\end{itemize}}

\newcommand{\bea}{\begin{align}}
\newcommand{\eea}{\end{align}}

\def\aap{A\&A}%
\def\apj{ApJ}%
\def\azh{AZh}%
\def\jcap{J. Cosmology Astropart. Phys.}%
\def\mnras{MNRAS}%
\def\prd{Phys.~Rev.~D}%
\def\prl{Phys.~Rev.~Lett.}%

\definecolor{myred}{RGB}{102,0,0}

\usepackage[pdftex,
            breaklinks=true,%
            colorlinks=true,%
            linkcolor=myred,
            urlcolor=blue,
            citecolor=blue,
            pdfauthor={Genolini et al.},%
            pdftitle={Neutron star - PBH interaction}%
           ]{hyperref}

\graphicspath{{./}{Figs/}}

\begin{document}
\preprint{LAPTH-028/20}
\title{Revisiting primordial black holes capture into neutron stars} 
\author{Y. G\'enolini}
\email{yoann.genolini@nbi.ku.dk}
\affiliation{Niels Bohr International Academy \& Discovery Center, Niels Bohr Institute, University of Copenhagen, Blegdamsvej 17, DK-2100 Copenhagen, Denmark}
\author{P. D. Serpico}
\email{serpico@lapth.cnrs.fr}
\affiliation{Univ. Grenoble Alpes, Univ. Savoie Mont Blanc, CNRS, LAPTh, F-74940 Annecy, France}
\author{P. Tinyakov}
\email{petr.tiniakov@ulb.ac.be}
\affiliation{Service de Physique Th\'eorique, Universit\'e Libre de Bruxelles, Boulevard du Triomphe, CP225, 1050 Brussels, Belgium}

\date{\today}

\preprint{}

\begin{abstract} 
Primordial black holes (PBH), produced through a variety of processes in the early universe, could fill galactic halos accounting for a fraction or the totality of the dark matter. 
In particular, PBH with sub-stellar masses could be captured by stars, start to swallow their material, and eventually turn them into BH, hence originating a peculiar new type of stellar catastrophic event. 
Here  we revisit this process in the most favorable case of PBH capture by neutron stars. We detail a number of novel features in the capture phase, during the settling  within the star and mass growth of the PBH, and illustrate some phenomenological consequences.  In particular, we point out that in the subsonic regime the PBH drag takes the form of a Bondi accretion. As a result, the onset of the final {\it transmutation} of the NS into a stellar sized BH is expected with the PBH seed in slight off-center position. 
We also compute the gravitational wave energy-loss and signals associated to different phases of the PBH-stellar interaction. In particular, the emission associated to the accretion phase is periodic with a few kHz frequency; in the rare case of a nearby Galactic event and for light PBH, it  could constitute a warning of the forthcoming transmutation. 
\end{abstract}
\maketitle

\section{Introduction}
\label{sec:intro}
Despite decades of tremendous experimental and theoretical efforts, the nature of dark matter (DM) is still elusive. Among the different possibilities, the idea that small collapsed structures, or primordial black holes (PBH) could account for all or part of the lacking mass is half-century old \cite{1966AZh....43..758Z,1971MNRAS.152...75H,Chapline:1975ojl}. While constraints on this hypothesis are strengthening for PBH with large masses, for relatively low PBH masses an interesting mass window $[10^{-16},10^{-10}]M_\odot$ is still presently unconstrained by lensing observation~\cite{2018JCAP...12..005K, 2019NatAs...3..524N, 2019arXiv190605950M}.

It is widely thought that interactions of PBH with compact stars are a promising avenue to shed light in that mass range, possibly via high-energy signatures. For example, it was proposed that a PBH crossing a white dwarf would trigger thermonuclear reaction of heavy elements and cause a runaway explosion leading to Type Ia supernova \cite{2015PhRvD..92f3007G}. In the case of a NS, the dense neutron medium is favourable for capture. Once trapped, the PBH grows and eventually swallows its host, transmuting it into a black hole (BH). The observation of old NS in DM-rich environments have already been used to set constraints on the PBH content of DM~\cite{2013PhRvD..87l3524C}. Furthermore, the transmutation process could lead to  signatures in various messengers and wavelengths, such as radio burst \cite{2015MNRAS.450L..71F,2018ApJ...868...17A}, kilonovae \cite{2018PhRvD..97e5016B}, positrons~\cite{Takhistov:2017nmt}, gamma-ray burst \cite{Takhistov:2017nmt,Chirenti:2019sxw}, and gravitational waves \cite{Takhistov:2017bpt,2018ApJ...868...17A,2016PhRvD..93b3508K,2007CQGra..24S.187B}. These different scenarios crucially depend on the dynamics of the seed BH growth, which ultimately impacts the amount of matter and energy expelled in this cataclysmic event. It was argued in \cite{2014PhRvD..90d3512K} that at early stage, the trapped PBH would smoothly accrete the NS material, evacuating the angular momentum through viscous dissipation. The final stages of the collapse on which the observational signatures critically depends on are however strongly uncertain. While realistic simulations coupling magneto-hydrodynamic and general relativity are probably the unique way to investigate the final signatures (see e.g. \cite{2019arXiv190907968E}), the details of the interactions of the PBH with the NS dense medium is of prime importance to assess the capture rate and set the initial conditions of the simulations. In this paper we revisit and refine the different interaction mechanisms, with a main focus on the capture and the post-capture dynamics and their observational consequences, as well as the role of gravitational wave (GW) emissions. 

The paper is organized as follows: In Sec.~\ref{sec:int_mech} we review the different energy-loss processes and discuss their velocity dependence. This includes, in Sec.~\ref{sec:int_gw}, the process due to gravitational wave (GW) losses, considered for the first time in this context. 
In Sec.~\ref{sec:capture}, we discuss the relative relevance of the different mechanisms for the capture, also assessing the role played by GW emission. In Sec.~\ref{eq:postcapture} we discuss the post-capture dynamics, which is remarkably simple and amenable to a description in terms of an adiabatic invariant (Eq.~(\ref{eq:conservation})) which leads us to establish a prescription for future numerical simulations, see Eq.~(\ref{initcond}). In Sec.~\ref{sec:signatures} we discuss a number of phenomenological consequences, with particular emphasis on the GW signatures of both the encounter and the post-capture dynamics, for individual events as well as the stochastic background.
Finally, in Sec.~\ref{concl} we discuss our results and conclude. 

In what follows we assume PBH masses $10^{-17}\,M_\odot \ll m\ll M_\odot$, so that: i) we can neglect Hawking radiation (and associated mass evaporation) in the whole evolution of the PBH, which  thus behave, in the absence of accretion, as stable objects. ii) The mass and size of the PBH is negligible with respect to the NS mass and size. This is anyway a very interesting mass range, where current upper limits~\cite{Carr:2020gox} on the fraction of DM in the form of PBH, $f_{\rm PBH}$, are typically not better than 1\%, and often closer to the 10\% level. In the range  $[10^{-16},10^{-10}]M_\odot$, bounds are absent or dependent on questionable assumptions, and PBH may also constitute the totality of the DM.

\section{Interaction mechanisms}\label{sec:int_mech}
By passing through (even close to) a NS, a PBH experiences several drag forces. While most of them have already been discussed in the literature, hereafter we review them, with a main focus on their velocity dependence. However, the content of Sec.~\ref{sec:int_coll} and especially the treatment of GW energy-losses (Sec.~\ref{sec:int_gw}) are novel  considerations in the context of the problem at hand. 
Note that, if the drag force ${\bold F}$ is known, the energy-loss can be promptly computed as the work of the force along the trajectory ${\cal C}$:
\beq
|\Delta E|= \int_{\cal C} {\bold F}\cdot {\rm d}{\bold l}\;.
\eeq

\subsection{Dynamical friction in a collisionless medium}\label{sec:int_dynfric}
As a PBH of mass $m$ passes through a collisionless medium, the gravitational pull from the wake of the PBH slows it down. 
This force  is called dynamical friction, and can be accounted for by the following formula \cite{RevModPhys.21.383,1987gady.book.....B}: 
\beq
{\bf F}_{\rm dyn}=-4\pi G^2 m^2\rho \ln{\Lambda_{\rm dyn} }(v) \frac{\boldsymbol{v} }{v^3}\;,\label{eq:df}
\eeq
where $G$ is Netwon's gravitational constant, $\rho$ the density of the medium and $\ln{\Lambda}_{\rm dyn}$ the so-called Coulomb logarithm, which depends on the ratio of extreme impact parameters. Notice that in our case, following \cite{2013PhRvD..87l3524C}, the Coulomb logarithm does depend on the velocity to account for the degenerate nature of the neutron fluid: neutrons contributing to the drag force are actually those for which the momentum transferred in the gravitational scattering is sufficient to extract them from the Fermi sea. The Coulomb logarithm writes~\cite{2013PhRvD..87l3524C}
\beq
\ln \Lambda_{\rm dyn} (v)= v^4 \gamma^2 \frac{2}{R_g^2}\int_{d_{\rm crit }}^{d_{\rm max}}{\rm d} x \,x (1-\cos \varphi(x))\;,\label{eq:ln_dyn}
\eeq
where $R_g=2\,G\,m$ is the Schwarzschild radius of the PBH,  $\varphi$ is the deviation angle of the neutron scattered by the PBH, $v$ its speed in the PBH reference frame and $\gamma$ is the Lorentz factor.  Below the critical PBH-neutron impact parameter $d_{\rm crit }$, the neutrons are accreted onto the PBH, while $d_{\rm max}$ is set by the requirement  that the scattered neutron must find an energy level  not already occupied by another neutron, or simply be ejected from the Fermi sea. Since the typical chemical potential of neutrons in a NS is $\mu_F \approx 0.3$~GeV, the effect of the degenerate matter reduces $\ln{ \Lambda}_{\rm dyn}$ by a factor $\approx 10$ at a speed $v = 0.8$. In summary, the typical energy-losses scales as:
\beq\label{Edyn}
|\Delta E|_{\rm dyn} \sim \frac{R_g^2 M_\star}{R_\star^2} \frac{\ln \Lambda}{ v_\star^2}\;,
\eeq
with $R_\star$ and $M_\star$ the stellar radius and mass, respectively, $v_\star^2=GM_\star/R_\star$ the typical PBH velocity, and the numerical pre-factor in Eq.~(\ref{Edyn}) is determined by the actual integration along the trajectory.

\subsection{Dynamical friction in a collisional medium}\label{sec:int_coll}
Eqs.~(\ref{eq:df},\ref{eq:ln_dyn}) strictly apply to a collisionless medium. This clearly may not be the case for the strongly interacting neutron fluid. However, the results must still be correct if the gravitational interaction timescale is much shorter than the causal time for the neutron-neutron interaction, set by the sound speed of the medium, $c_s$. We expect thus that Eqs.~(\ref{eq:df},\ref{eq:ln_dyn}) are valid for a PBH moving at {\it supersonic speed} in the neutron fluid.  This is confirmed by the study of friction in a collisional medium developed in \cite{1999ApJ...513..252O}.  Its results can be summarized as follows: i) At ${\cal M}\equiv v/c_s\gtrsim 2$, the collisionless result is reproduced. ii) At $1\lesssim{\cal M}\lesssim 2$,  the friction force is resonantly enhanced. iii) For speeds smaller than the sound speed, for a transient of 
the order of $R_\star/c_s$, the PBH feels a force given by Eq.~(\ref{eq:df}) with $\ln \Lambda_{\rm dyn}$ replaced by 
\beq
\ln{\Lambda}_{\rm coll}=\frac{1}{2}\ln\left( \frac{1+{\cal M}}{1- {\cal M}}\right)-{\cal M}\simeq \frac{{\cal M}^3}{3} \:{\rm for}\: {\cal M}\ll 1\;.\label{eq:ln_dyn_sub}
\eeq
Eventually, however, the friction force tends to zero when the system settles closer and closer to the steady state limit.

\subsection{Accretion}\label{sec:int_accr}
From Newton's laws, the accretion of matter with a rate $\dot m$ and zero momentum causes a drag force in the opposite direction of the motion:
\beq
{\bf F}_{\rm acc}= -\;\dot m \;\boldsymbol{v}\;.\label{eq:facc}
\eeq
In the supersonic regime, as argued in \cite{2013PhRvD..87b3507C}, this force can be written as Eq.~(\ref{eq:df}), with $\ln \Lambda_{\rm dyn}$ replaced by 
\beq
\ln \Lambda_{\rm acc}(v)= v^4 \gamma^2 \frac{d_{\rm crit}^2}{R_g^2}\;, \label{eq:ln_acc_p}
\eeq 
with the same notation as used in Eq.~(\ref{eq:ln_dyn}).

In the subsonic regime, an analytical theory only exists rigorously for a body accreting at rest, assumed to apply for very small speeds $v \ll c$. In this case, 
the accretion rate $\dot m$ tends to the spherical Bondi accretion rate \cite{1952MNRAS.112..195B}
\beq
 \dot m=\frac{{\rm d}m}{{\rm d
 }t} =\frac{4\pi\,\lambda\, \rho \,G^2 m^2}{c_s^3}\;,\label{eq:bondi}
\eeq
with $\lambda$ depending on the medium properties, equal to  0.707 for a polytropic equation of state with index $\Gamma=4/3$~\cite{Kouvaris:2013kra}. Although some expressions valid for finite $v$, such as
\beq
 \dot m=\frac{{\rm d}m}{{\rm d
 }t} =\frac{4\pi\,\lambda\, \rho \,G^2 m^2}{(v^2 + c_s^2)^{3/2}}\;,\label{eq:bondi_v}
\eeq
have been proposed already in~\cite{1952MNRAS.112..195B} and roughly confirmed by simulations \cite{1985MNRAS.217..367S}, the correction to Eq.~(\ref{eq:bondi}) is expected to be small in the deeply sub-sonic regime of major interest in our paper, and we will neglect it in the following. 
It is worth noting that the accretion force can be written as Eq.~(\ref{eq:df}), with $\ln \Lambda_{\rm dyn}$ replaced by 
\beq
\ln \Lambda_{\rm sub}(v)=\lambda \,{\cal M}^3\,\label{eq:ln_acc_m}.
\eeq

\subsection{Surface waves}\label{sec:int_surf}

A PBH crossing the neutron star excites waves on its surface.  These surface waves  --- essentially, tidal deformations of the NS --- are different from the sound waves; in particular, they have a different dispersion relation. For this reason we discuss them separately.

The total energy dissipated by the BH into production of surface waves has been estimated in Ref.~\cite{2014PhRvD..90j3522D} by making use of a simple analytical model, an incompressible fluid in a uniform gravitational field. The result, Eq.~(13) of Ref.~\cite{2014PhRvD..90j3522D}, up to a numerical coefficient leads to the following  estimate for the energy-loss in a NS
\beq
|\Delta E|_{\rm surf}  \sim \frac{G\,m^2}{R_\star}\;.\label{eq:swel}
\eeq
Keeping in mind that $G\,M_\star/R_\star\gtrsim 0.2$ for a NS, this is parametrically similar to the energy-loss due to the dynamical friction, Eq.~(\ref{Edyn}), however {\em without the enhancement associated to  the Coulomb logarithm}. 

It is easy to understand this result intuitively in terms of the dynamical friction calculation. In the case of an infinite medium, the energy-loss from dynamical friction gets contributions from all distances, hence the logarithmic divergence of the sum. When applied to the star, the star radius $R_\star$ imposes a cutoff, which essentially means retaining only leading-log contributions. Changing the shape of the star would change the subleading constant  term. Similar features are shared by surface waves.

Such arguments suggest that the contribution to the energy-loss due to the {\it tidal deformations} induced by PBH  passing near NS {\em (without actually crossing its surface)} is also subleading.  One may view this process as dynamical friction in an infinite medium, in which contributions of all volumes are switched off except the one actually occupied by the star. The Coulomb log now becomes $\log{[r_{\rm min} / (r_{\rm min} - R_\star )]}$, $r_{\rm min}$ being the periastron of the PBH orbit. Clearly this logarithm becomes subleading to the original $\log{(R_\star/R_g)}$ way before $r_{\rm min} - R_\star$ becomes comparable to $R_\star$. The fact that only the part of the volume actually occupied by the star is filled with matter reduces further this contribution, resulting in the suppression by  some power of the ratio $R_\star/r_{\rm min}$. The actual calculation yields a series of the type
\beq
|\Delta E|_{\rm tidal}  \sim \frac{G m^2}{R_\star} \sum_{l=2}^\infty
\left( {R_\star\over r_{\rm min}} \right)^{2\ell+2} T_\ell,
\;.\label{eq:tidal}
\eeq
where each term is suppressed by $[R_\star/r_{\rm min}]^{2\ell+2}$ \cite{1977ApJ...213..183P}, with  $\ell$  the multipole number of the tidal deformation, and $T_\ell$ are dimensionless coefficients $\lesssim {\cal O}(1)$ that depend on the star properties and $R_\star/ r_{\rm min}$. Clearly, this expression is greatly suppressed for
$R_\star/ r_{\rm min}\ll 1$, while for $R_\star/ r_{\rm min} \to 1 $ it must turn into the energy-loss due to surface wave emission. 

\subsection{Gravitational waves}\label{sec:int_gw}
The encounter of the relativistically moving PBH with the compact star produces gravitational waves (GW), which is the only energy-loss mechanism we take into account in some detail which does not strictly require contact to be operational. For hyperbolic encounters, analytical calculations for generic configurations  have been performed in~\cite{Capozziello:2008ra} and~\cite{DeVittori:2012da}.  In the context of pairs of PBH of stellar masses, this process has been considered in~\cite{2017PDU....18..123G,2018PDU....21...61G}. Here, we illustrate the results of these calculations, applying them to the case of interest, but also {\it generalize the calculation to the case of a PBH crossing the NS}, with the latter  reported in Sec.~\ref{sec:gw}.

The power in GW is 
\begin{equation}
\frac{{\rm d}E}{{\rm d}t}=\frac{G}{5\,c^5}\langle \dddot{Q}_{ij} \dddot{Q}_{ij}\rangle\,,\label{PowGW}
\end{equation}
where we introduced the quadrupole 
\begin{equation}
Q_{ij}\equiv\int \rho(\boldsymbol{r}) (r_ir_j-\frac{1}{3}  r^2 \delta_{ij})\,d^3 r\;.
\end{equation}
We actually re-express the energy-loss in terms of the distance $d$ to the source and the  gravitational strain $h_{ij}$, defined as:
\beq
h_{ij}=\frac{2}{c^4}\frac{G}{d}\ddot{Q}_{ij}\,.
\eeq

We compute the typical gravitational strain in cartesian coordinates, expressing it as
\beq
h_{0}=(h_{xx}^2+h_{yy}^2+2h_{xy}^2)^{1/2}=\frac{R_g v_i^2}{c^4\,d}\, g(\phi,e)\;,\label{eq:gw_general}
\eeq
with $g(\phi,e)$ a complicated function that depends on the eccentricity $e$ and the phase angle $\phi=\phi(t)$. Its expression can be found in Appendix~\ref{app:pbh_motion}, where we give the detailed description of the orbit of the PBH depending on the initial PBH speed $v_i$ and impact parameter $b$.

The contribution to the energy radiated in GW can be split in two pieces, respectively accounting for the motion inside and outside the NS,
\beq
|\Delta E|_{\rm gw}= \Delta E_{\rm gw}^{\rm in} + \Delta E_{\rm gw}^{\rm out} \; \label{eq:gwel}.
\eeq
For the purpose of the capture, in the case when the PBH crosses the NS surface $\Delta E_{\rm gw}$ (and hence $\Delta E^{\rm in}_{\rm gw}$) is never important compared to the other contributions previously described. On the other hand, for  larger impact parameters $\Delta E_{\rm gw}^{\rm out}$ may be relevant. One can borrow directly Eq.~(3.13) from  \cite{DeVittori:2012da}. Rewriting it in terms of the eccentricity $e$ and the periastron distance $p(e)$, and denoting $M\equiv m+M_\star$, we have 
 \begin{equation}
\Delta E_{\rm gw}=\frac{8}{15}\frac{m^2M_*^2}{M^3}v_i^7\frac{p(e)}{(e-1)^{7/2}}\,.  \label{DeltaEGW}
\end{equation}
Here the eccentricity $e$ of the orbit is related to the relevant independent parameters $b,v_i, m, M_\star$ via
\begin{equation}
e=\sqrt{1+\frac{b^2}{a^2}}=\sqrt{1+\frac{b^2v_i^4}{G^2M^2}}
\,,\label{eccentricity}
\end{equation}
where we also introduced the semi-major axis $a=\sqrt{G^2M^2/v_i^4}$. For completeness, the function $p(e)$ is given by
\[
p(e) = (e+1)^{-7/2} \Biggl\{
\arccos\left(-\frac{1}{e}\right)\left(24+73\,e^2+\frac{37}{4}e^4\right)
\]
\begin{equation}
+\frac{\sqrt{e^2-1}}{12}(602+673\,e^2)\Biggr\} \,\nonumber.
\end{equation}
It is straightforward to derive the scaling $\Delta E_{\rm gw}\propto v_i^{-7}$ in the regime of physical interest here ($e\approx 1$), 
which suggests a growing relative importance of this energy-loss channel for low velocity dispersion systems.

\section{Capture}
\label{sec:capture}

A PBH gets captured by a NS, i.e. becomes gravitationally bound to it, if it loses enough kinetic energy so that its total energy becomes negative. The capture condition of a PBH of mass $m$ thus writes:
\beq
 |\Delta E|>E_i=\frac{1}{2}m v_i^2 \;, \label{eq:cap_condition}
\eeq
with $v_i$ the PBH velocity at infinity and $\Delta E$ the energy-losses coming from the different interaction mechanisms reviewed. Based on the previous discussion, it is important to assess if the PBH is moving supersonically or subsonically when it interacts. We anticipate that, for a broad range of NS models, the first interaction (hence the possible capture) of the PBH with the NS material happens at supersonic or transonic velocities, i.e. ${\cal M}\gtrsim 1$. To reach this conclusion, we took several benchmark models from the literature \cite{2013A&A...560A..48P} (BSK-20 and BSK-21) spanning equations of state of varying stiffness, with a low and a high mass model which are in agreement with the LIGO/VIRGO data from a binary neutron star merger~\cite{Abbott:2018exr}. For the sake of clarity we define the typical velocity $v_\star$, associated angular velocity  $\omega_\star$, frequency $f_\star$, period $T_\star$,
\beq
v_\star=\sqrt{\frac{GM_\star}{R_\star}}, \: \: \omega_\star=\frac{v_\star}{R_\star}, \: \: f_\star=\frac{1}{T_\star}=\frac{\omega_\star}{2\pi}\:,
\eeq
that we will use throughout our calculations. 
These scales are summarized in Tab.~\ref{tab:numbers} for the profiles considered; we also report the sound speed ($c_s$) and the chemical potential of neutrons ($\mu_n$) in the NS core. Unless written otherwise, the numerical estimates given in the paper are based on the values of the BSK-20-1 NS model, and assume that the NS is an homogeneous sphere of matter.

\begin{table}[h!]
	\begin{center}
	\begin{tabular}{l c c c c}
	\hline\hline
	Model & BSK-20-1 & BSK-20-2 & BSK 21-1 & BSK 21-2  \\
	\hline 
	Radius $R_\star$ [km] &  11.6 & 10.7 & 12.5 & 12.0\\
    Mass $M_\star$ [$\rm M_{\odot}$] & 1.52 & 2.12 & 1.54 & 2.11\\
    $v_\star$ [$c$] & 0.44 & 0.54 & 0.43 & 0.50 \\
    $f_\star=1/T_\star$ [kHz] & 1.8 & 2.4 & 1.6 & 2.0\\ 
    $c_s$ (core) [$c$] & 0.68 & 0.97 & 0.64 & 0.81  \\
    $\mu_n$ (core) [GeV] & 0.27 & 0.81 & 0.24 & 0.51  \\
	\hline
	\end{tabular}
	\caption{Relevant parameters for the benchmark NS models considered.}
    \label{tab:numbers}
	\end{center}
\end{table}

\begin{figure}[h!]
    \centering
    \includegraphics[width=\columnwidth]{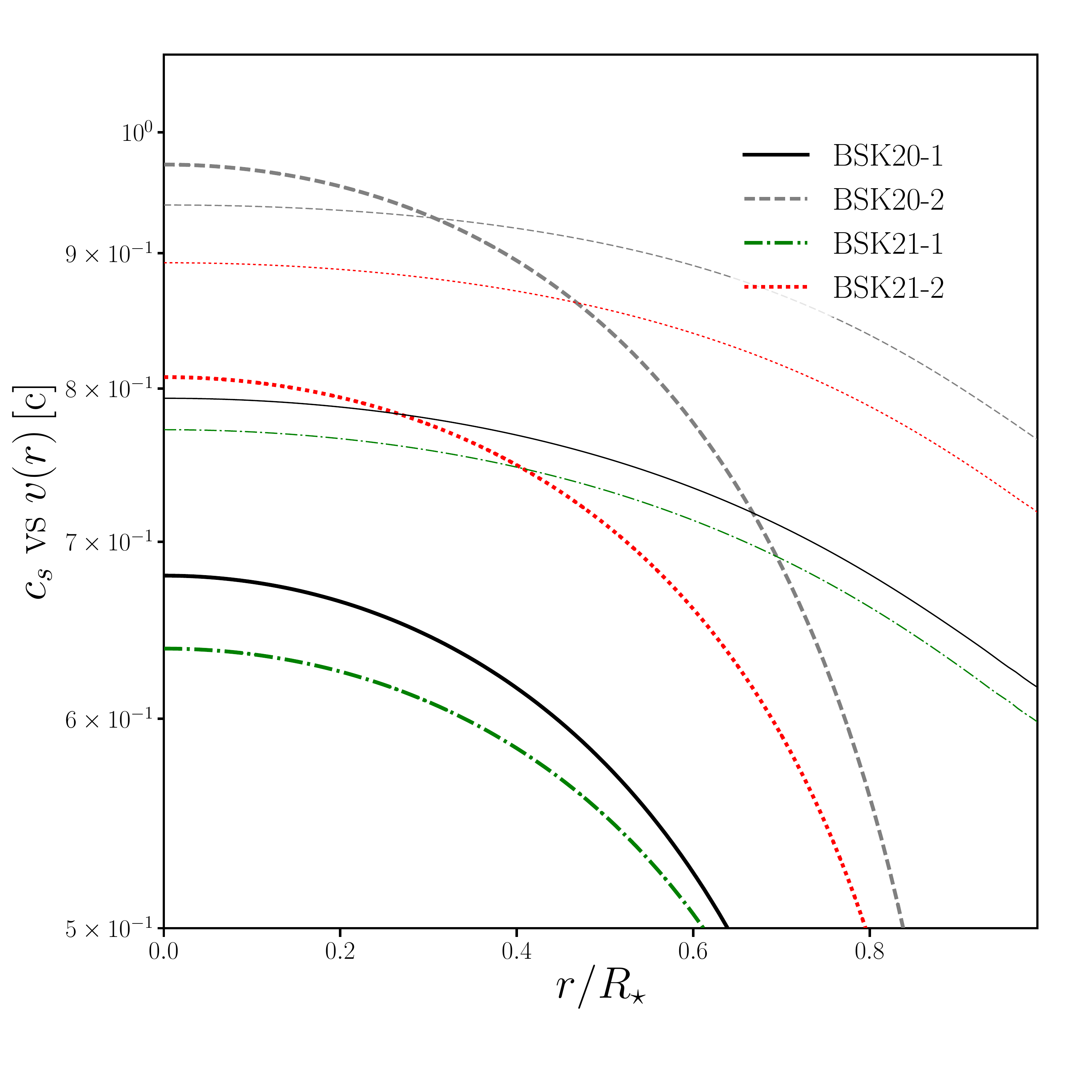}
    \caption{Sound speed (thick line) and PBH speed (thin line) as a function of the radius for the NS reference profiles considered.\label{fig:sound_speed}}
\end{figure}

\begin{figure}[h!]
    \centering
    \includegraphics[width=\columnwidth]{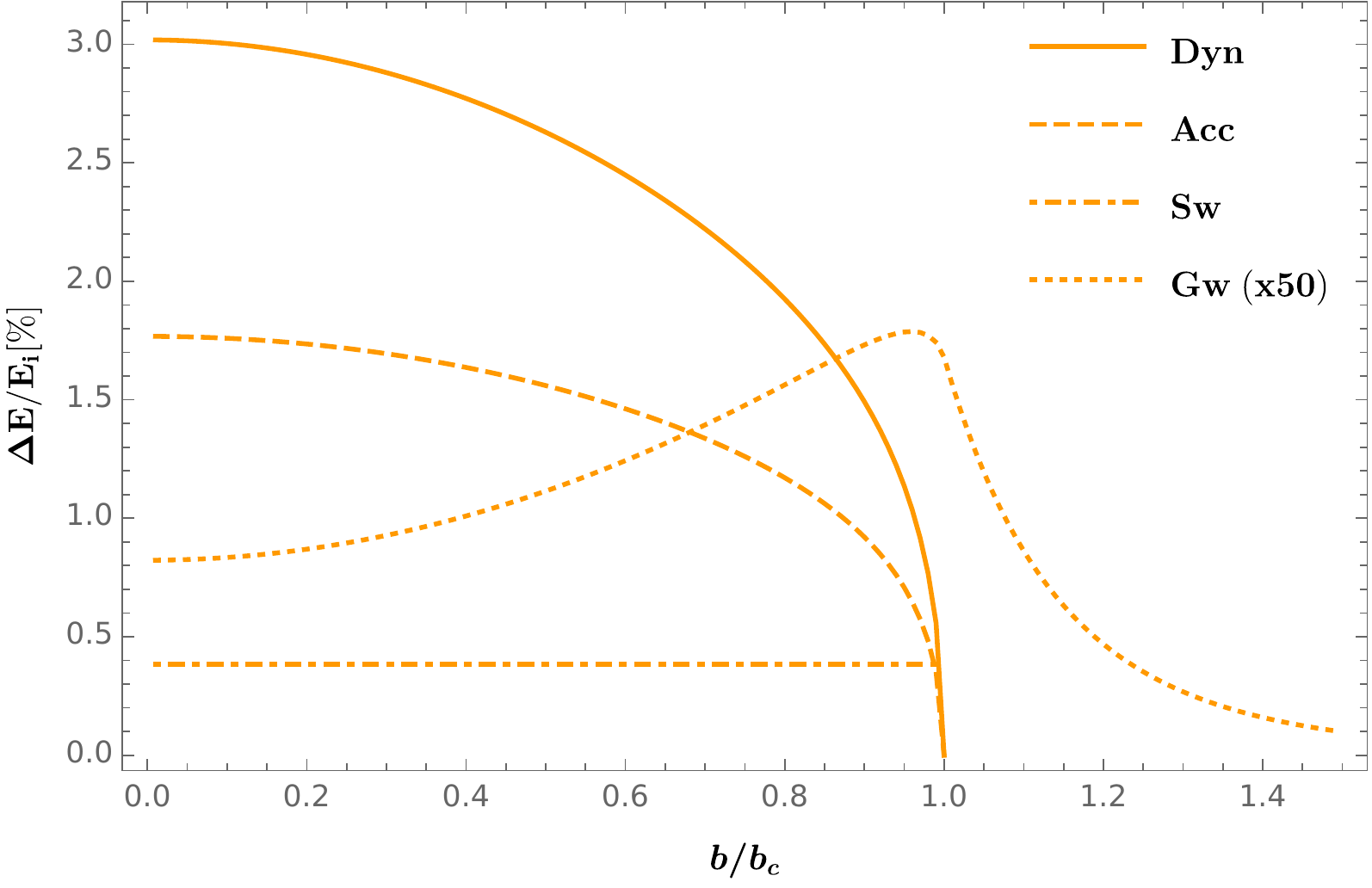}
    \caption{Comparison of the contribution of the different processes of energy-losses to capture as a function of the impact parameter ratio $b/b_c$, for the benchmark $v_i=10^{-3}$. 
    \label{fig:energy_losses}}
\end{figure}

To discuss if capture happens in the subsonic or supersonic regime, one should compare the speed of the PBH travelling through the NS with the sound speed of the NS medium along its trajectory. The trajectory of the PBH in the NS is presented in Appendix~\ref{app:pbh_motion}.
In the limit $v_i \ll v_\star$, the arrival speed of a PBH as a function of $r$ is given by the following expression in terms of the gravitational potential $\Phi$,
\beq
v(r)=\sqrt{1-e^{2(\Phi(\infty)-\Phi(r))}}\simeq v_\star\,\sqrt{3-r^2/R_\star^2}\;, \label{eq:vesc}
\eeq
where the first expression at the RHS takes into account GR effects, and the second approximate equality holds in the Newtonian limit, being accurate to within 5\%. In Fig.~\ref{fig:sound_speed} we show the sound speed of the chosen NS benchmark models as a function of $r$ (thick lines), as well as the velocity Eq.~(\ref{eq:vesc}) (thin lines of corresponding style and color). For all models except BSK-20-2, the PBH speed is always larger than the sound speed  at any given $r$. For BSK-20-2, the velocity can drop slightly below (few percent) the speed of sound if the PBH enters within the inner third of the star.
Using the expressions valid in the {\it supersonic regime} for the mechanisms summarized in Sec.~\ref{sec:int_mech}, we compute the different energy-losses as a function of the impact parameter. To this goal, we define $b_c$ as the critical impact parameter, such that a PBH having $b=b_c$ will eventually graze the NS of radius $R_\star$,  reaching in its orbit a minimal distance from the center $r_{\rm min}=R_\star$. In terms of the initial velocity $v_i$, one has:
\beq
b_c=R_\star \sqrt{1+2\frac{v_\star^2}{v_i^2}}
\;.\label{eq:bc}
\eeq
As an example, in the model BSK-20-1 we obtain $\tilde{b}_c\equiv b_c/R_\star\approx 624$. Our results are reported in Fig.~\ref{fig:energy_losses} for the initial velocity value  $v_i=10^{-3}$. 
It is clear that the dominant process for capture is dynamical friction whatever the impact parameter $b<b_c$. For all but the GW term, the vertical axis scales roughly as $1/E_i\propto (10^{-3}/v_i)^2$. For initial velocities $v_i<2\times 10^{-4}$, GW capture for $b>b_c$ becomes important. Note that, had we used expressions for the transonic regime, the impact of the dynamical friction would have been enhanced thanks to the resonant effect mentioned in Sec.~\ref{sec:int_coll}, while the accretion mechanism would have been comparatively suppressed. We conclude, consistently with the common lore, that dynamical friction is the dominant mechanism for energy-loss by the PBH passing through the NS. However, it is not always a dominant {\it capture} mechanism, as we will argue in Sec.~\ref{sec:signatures}.

A naive look at Fig.~\ref{fig:energy_losses} would suggest no capture for the typical velocity dispersion in the Milky Way halo. However, this would be incorrect, given the broad
distribution of velocities. To be more quantitative, we assume that the PBHs follow a Maxwellian distribution in velocities with the dispersion $\bar{v}$,
\begin{equation}
{\rm d}^3n=n_\text{PBH}\left(\frac{3}{2\pi \bar{v}^2}\right)^{3/2} 
\exp\left\{\frac{-3v^2}{2\bar{v}^2}\right\} {\rm d}^3v,
\end{equation}
where $n_\text{PBH}=\rho_\text{PBH}/m$, $\rho_\text{BH}$ is the density of PBHs at the star location, and $m$ their mass. It can be expressed in terms of the local DM density $\rho_\text{DM}$ as follows, 
\begin{equation}
\rho_\text{PBH} = f_{\rm PBH} \rho_\text{DM},
\label{eq:rhoBH}
\end{equation}
with $ f_{\rm PBH}$ that can observationally attain values as large as 1 for $10^{-16}\,M_\odot \lesssim m\lesssim 10^{-10}\,M_\odot$, while being limited to $f_{\rm PBH}\lesssim{\cal O}$(0.01-0.1) for $10^{-10}\,M_\odot\lesssim m\lesssim 0.1\,M_\odot$. In the following we always assume that $\bar{v}\ll v_\star$.

\begin{figure}[h!]
    \centering
    \includegraphics[width=\columnwidth]{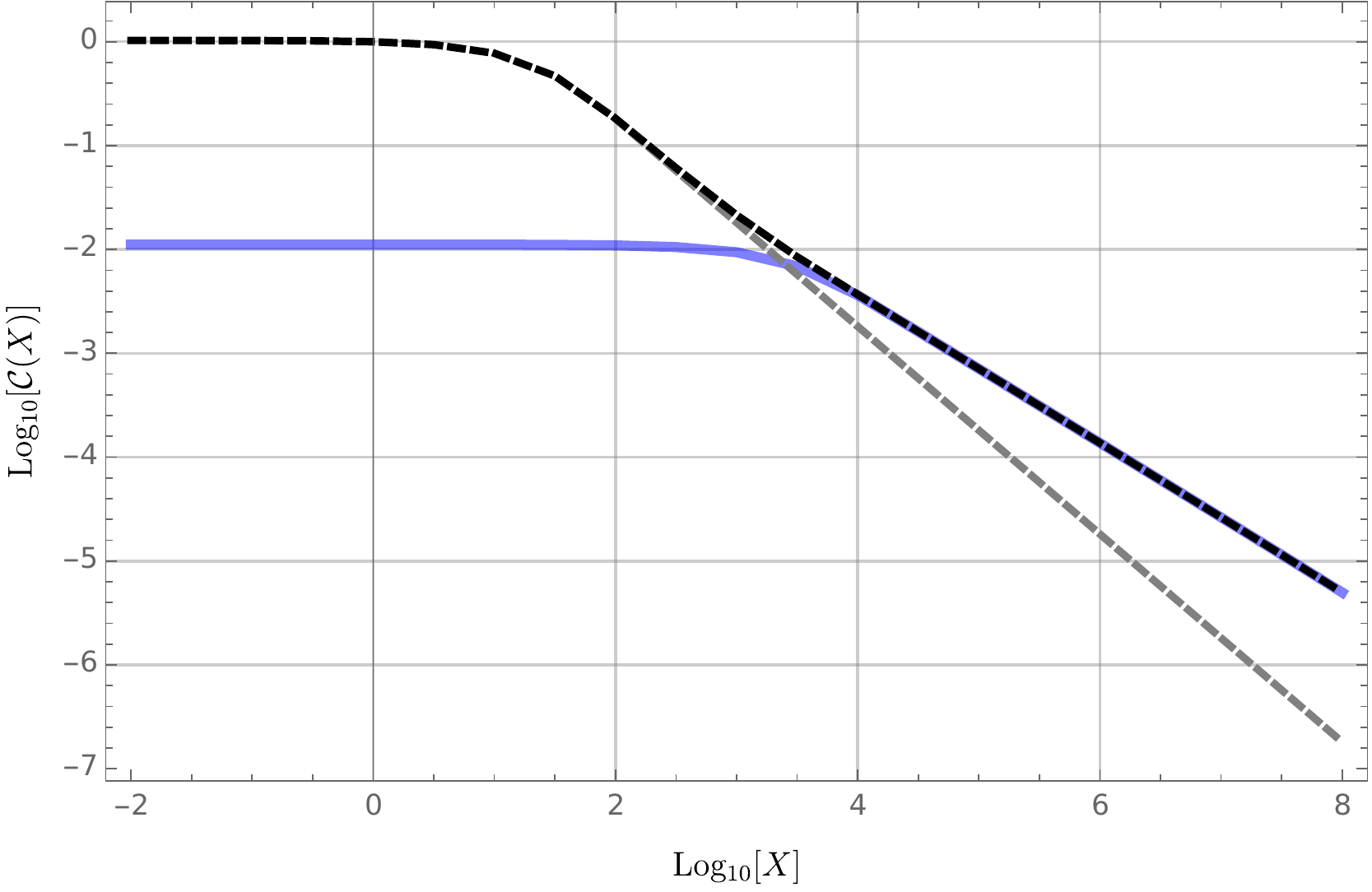}
    \caption{Evolution of the function ${\cal C}$ (black-dashed line) of Eq.~(\ref{eq:cap_rate}) as a function of $X$ defined in Eq.~(\ref{eq:defX}). The sole contribution of GW capture is displayed in blue and the difference with the total is shown with a dashed gray line.\label{fig:CX_scaling}}
\end{figure}

The rate of NS-PBH encounter leading to capture is:
\beq
{\cal G}_{\star}=\int \frac{{\rm d}^3 n}{{\rm d}v^3}\; {\cal S}(v) \; v\; {\rm d}^3v\;,
\eeq
where ${\cal S}(v)=\pi\; b_{\cal G}^2$ is the effective cross-section of the star which leads to capture\footnote{Note that a more precise GR treatment accounting for the Schwarzschild metric of the NS would lead to an enhanced capture rate by a factor $1/(1-R_\star^s/R_\star)\approx 1.6$~\cite{2008PhRvD..77b3006K}, where  $R_\star^s$ is the Schwarzschild radius of the NS.}. This is defined by the condition in Eq.~\ref{eq:cap_condition}: In practice, $b_{\cal G}(v)$ is the {\it largest} $b$ solving the implicit equation $\Delta E(b,v,m,M_\star)=m v^2/2$, where $\Delta E$ includes all energy-losses.  
Considering all the processes discussed above, and using the typical PBH velocity dispersion $\bar{v}=10^{-3}$, we find numerically:
\beq
{\cal G}_{\star}\simeq 2.1\times 10^{-17} \; \left( \frac{\rho_\text{PBH}}{\rm GeV\,cm^{-3}}\right) \left( \frac{10^{-3}}{\bar v}\right)^3 {\cal C}\left[X \right] \rm yr^{-1}\;,\label{eq:cap_rate}
\eeq
with,
\beq
X=X(m,{\bar v})\equiv \left( \frac{m}{10^{25}\rm g}\right)\left( \frac{10^{-3}}{\bar v}\right)^2\;\label{eq:defX}\,,
\eeq
and 
the function ${\cal C} [X]$ is displayed with a dashed-black line in Fig.~\ref{fig:CX_scaling}.  Because of the form of ${\cal S}(v)$, the dependence on ${\bar{v}}$ and $m$ is not trivial. The contribution of GW to ${\cal C} [X]$ is shown with a blue line in Fig.~\ref{fig:CX_scaling}, whereas  the dashed-gray curve displays the behaviour of $\cal C$ without accounting for gravitational capture. For $X<10$, ${\cal C} [X]$ is constant; for $10<X<10^3$ it declines as $X^{-1}$; when $X>10^3$, the decline follows the milder behaviour $\propto X^{-5/7}$, because capture by GW emission kicks in and dominates at large impact parameters. Note that, although at large $X$ the capture is suppressed, the GW capture becomes comparatively  more important. If fixing the mass at $10^{25}\rm g$, at $\bar{ v}=10^{-3}$, $10^{-4}$ and $10^{-5}$ the GW capture is responsible for 1.1\%, 6.0\% and 99\% of the captures, respectively. 

In obtaining the above results, we have considered only interactions between the PBH and an {\it isolated} NS. While a detailed account of {\it multi-body effects} goes beyond our goals, let us mention the current understanding of these processes. If the NS is in a tight binary, it has been shown in~\cite{2012PhRvL.109f1301B} that the capture can be enhanced by a factor up to 3-4, due to the energy-loss of the PBH (or any ``test particle'', for what matters) resulting from its gravitational scattering off the NS moving companion. More frequently, the PBH falling onto the NS will also experience tidal effects by the stellar clusters or even by the Galactic disk in which  the NS is embedded. For the Milky Way disk, this effect (which in general may either enhance or deplete the capture probability) has been estimated to become important in the capture process for $m\lesssim {\rm few}\times 10^{-13}\,M_\odot$~\cite{2019arXiv190605950M}.

For comparison, we also define the rate of encounters which involve interaction with matter, but does not always lead to capture:
\begin{align}
\Gamma_{\star}&=\int \frac{{\rm d}^3 n}{{\rm d}v^3}\; \pi b_c^2(v) \; v\; {\rm d}^3v=\nonumber\;\\
&\simeq 3.8\times 10^{-16} \; \left( \frac{\rho_\text{BH}}{1 \rm GeV\, cm^{-3}}\right)\left( \frac{10^{25}\rm g}{m}\right) \left( \frac{10^{-3}}{\bar v}\right)\rm yr^{-1}\;,\label{encontTOT}
\end{align}
where $b_c$ is defined in Eq.~(\ref{eq:bc}). 
Further comments on these capture rates and potentially observable consequences are reported in Sec.~\ref{sec:signatures}. 

\section{Post-capture} \label{eq:postcapture}

If the PBH first interaction leads to its capture, it starts orbiting on a bounded trajectory, typically with large eccentricity. We can distinguish two cases, according if it was captured with an interaction outside or inside the NS. In the former case, the GW emission makes the PBH to settle on a meta-stable elliptical orbit around the NS, for a timescale
\beq
t_{\rm settle}^{\rm GW}\simeq 16\; \left(\frac{m}{10^{22}\,{\rm g}}\right)^{-3/2} \left(\frac{b}{b_c} \right)^{21/2}\left(\frac{v_\star}{0.44}\right)^{-19} \rm Myr\;.\label{tgwsettl}
\eeq
This time is estimated from the coalescence time of high eccentricity binaries from Ref.~\cite{PhysRev.136.B1224}, taking the same periastron for the elliptical trajectory as the hyperbolic one along which the PBH is captured, and choosing a binding energy of order $\Delta E_{\rm gw}$. This timescale is quite sensitive to the values of $v_\star$; for the reference values chosen, it becomes shorter than the age of the Universe $t_U$ for $m>1.4\times 10^{20}\,{\rm g}\simeq 7\times10^{-14}\,M_\odot$. 

In the latter case, when the capture happens via an interaction inside the NS, or once the PBH drops inside the NS via GW losses, the energy-loss timescale is much shorter. The PBH mostly loses energy each time it passes through the NS, eventually settling on a fully contained orbit around the star center in a timescale~\cite{2013PhRvD..87l3524C}
\beq
t_{\rm settle}\lesssim 4\times 10^4\,\left(\frac{m}{10^{22}\,{\rm g}}\right)^{-3/2}\,{\rm yr}.\label{tsettl}
\eeq
This is shorter than $t_U$ for $m>2\times 10^{18}\,{\rm g}\simeq 10^{-15}\,M_\odot$.
While during the first passage the PBH crosses the NS with a supersonic velocity, at later stages when the orbit size becomes smaller than $r\lesssim R_\star c_s/v_\star$ the PBH motion becomes subsonic. 
From this moment onward one may neglect all the contributions to the drag force except the one due to accretion of ambient matter, so that
\beq
{\bf F}_{\rm drag} = - \dot m \boldsymbol{v} = 
-4\pi G^2 m^2\rho  \frac{\boldsymbol{v} }{c_s^3}\;\label{pcdrag}
\eeq
where $\boldsymbol{v}$ is the {\em relative} velocity of the PBH and the ambient matter. 
The equation of motion of the PBH in this case takes the following form:
\beq
\ddot{\boldsymbol{r}} +{\cal D}(t) 
\left[
\dot {\boldsymbol{r}} - \boldsymbol{\Omega}\times \boldsymbol{r} \right] + \omega_\star^2 {\boldsymbol{r}}=0\; \label{eq:motion},
\eeq
where the PBH position $\boldsymbol{r}$ is defined with respect to the star center,  ${\cal D}(t) = \dot m/m$ (cf. Eq.~(\ref{pcdrag})),  
$\omega_\star = \sqrt{4\pi G\rho/3} \sim 1.1\times 10^4\,$s$^{-1}$ is the angular velocity around the NS center, and we have included the possibility that the NS rotates with angular velocity $\boldsymbol{\Omega}$. 

This equation factorizes into three independent damped harmonic oscillator equations: one for the motion $r_3(t)$ parallel to 
$\boldsymbol{\Omega}$, and two equations for a co-rotating and counter-rotating modes $r_\pm (t)= r_1(t)\pm i r_2(t) $ in the plane orthogonal to $\boldsymbol{\Omega}$. The damping term in these equations is small,
\beq
{{\cal D}\over \omega_\star} \sim 2.8\times 10^{-12} \left( { m \over 10^{22} {\rm g}}\right)  \ll 1,
\label{eq:oscillation_cond}
\eeq
and slowly varying with time. The approximate solution is then written in the form
\beq
r_\pm \propto \exp \left\{ - {1\over 2} \left( 1\mp {\Omega\over\omega_\star}\right)
\ln m + i\omega_\star t\right\}.
\label{eq:oscillations-solution}
\eeq
The same solution with $\Omega=0$ is valid for $r_3(t)$. 

For most of the observed  NS the ratio $\Omega/\omega_\star$ is much smaller than 1, reaching about $1/4$ for the fastest millisecond pulsar. Thus, the correction due to NS rotation in Eq.~(\ref{eq:oscillations-solution}) can be neglected in most of the cases. Then all three solutions have the same behavior which implies 
\beq
m\;r^2\;=\; {\rm const.}\label{eq:conservation}
\eeq
Note that this (approximate) conservation law does not depend on the accretion regime, as long as ${\cal D}\ll \omega_\star$.  

Making use of this relation, one may readily estimate a typical displacement of the oscillating PBH form the star center at a time when its mass has grown to a fraction $f\ll 1$ of the star mass, $m=f M_\star$. Assuming initial mass $m_i$ and initial orbital radius $r_i\sim R_\star$, the final orbital radius is 
\beq
R_f=R_\star \sqrt{\frac{m_i}{f\,M_\star}}\,.\label{initcond}
\eeq

To conclude this section, let us estimate the time it would take a PBH of mass $m$ settled within the NS to accrete the whole star. For the rough estimate we approximate the star as a medium of constant density $\rho_\star=3M_\star/(4\pi R_\star^3)$. This is a reasonable approximation in the center of a NS, whose typical profile goes as $\rho_\star \propto (1-(r/R_\star)^2)^{1/2}$ \cite{2013A&A...560A..48P}. Assuming the Bondi accretion rate, Eq.~(\ref{eq:bondi}), 
we obtain
\beq
m(t)=\frac{m}{1-t/t_B}\,,
\eeq
where
\beq
t_{B}= \frac{ c_s^3\, R_\star^3}{3 \,G^2 \,M_\star\, m}\simeq 1\left( \frac{10^{22}\rm g}{m}\right) \rm yr
\eeq
is the typical time needed for the PBH to consume the whole NS. Actually, as discussed in Ref.~\cite{2014PhRvD..90d3512K}, the Bondi regime may fail before than the whole star is consumed and  $m(t)\simeq M_\star$, the reason being the angular momentum conservation. In this case the Bondi regime is probably replaced by Eddington-like accretion during the last stages, slightly prolonging the life of the star.

\section{Signatures}\label{sec:signatures}
The dynamics outlined in the previous sections  has a number of phenomenological consequences, which we discuss in this section. 

\subsection{GW bursts from typical PBH-NS encounters}\label{sec:gw}
In a hyperbolic encounter between two massive objects, a characteristic ``tear drop'' burst signal is emitted, according the  LIGO nomenclature~\cite{2017CQGra..34c4002P}. A similar signature for PBH has been considered in encounters between pairs of PBH in~\cite{2018PDU....21...61G,2017PDU....18..123G}.
Apart for the masses of the two bodies, the motion depends on the impact parameter $b$ and the initial speed $v_i$ or, equivalently, the eccentricity $e$ given by Eq.~(\ref{eccentricity}). 
The GW signal can then be computed as explained in Sec.~\ref{sec:int_gw}, provided that the orbital function $g(e,\phi(t))$ is known. 
In the limit of monochromatic emission and for $m\ll M_\star$ one can describe the strain due to a hyperbolic encounter as producing a typical GW burst of amplitude $h_c(b,v_i,d)$ and characteristic frequency $f_c(b,v_i)$ as in Refs.~\cite{2017PDU....18..123G, 2018PDU....21...61G} \footnote{Note that a factor $1/3$ is missing in their definition of $h$, given the expression they take for the quadrupole. }:
\begin{align}
h_c(b,v_i,d)&=\frac{2Gm}{3\; d\; c^2}\beta_i^2 \frac{2}{e-1}\sqrt{18(e+1)+5e^2}\\
f_c(b,v_i)&=\frac{1}{2\pi}\frac{v_i}{b}\frac{e+1}{e-1}\,,
\end{align}
where $d$ denotes the  distance of the observer from the encounter. For $b=b_c$, $v_i=10^{-3}$, $d=1\,$kpc and $m=10^{25}$g, typical values are $h_c\approx 4\times 10^{-25}$ and $f\approx 3\;$kHz. 

Note that these functions diverge for small impact parameter, i.e. $b \to 0$. Hence in the following,
we generalize the calculation to the case where the PBH passes within the NS. We consider a perturbative approach in which the GW emission is computed along the unperturbed trajectory. Outside the NS, both before entering and after exiting the star, the motion is hyperbolic with parameters determined as from  Eq.~(\ref{eccentricity}). Inside the NS, 
in the approximation of constant density, the gravitational potential is a harmonic potential. Hence, within the star, the PBH follows an elliptical orbit centered on the NS center, with  semi-minor axis $\alpha_-$ and semi-major axis $\alpha_+$ given by
\beq
\frac{\alpha_\pm}{R_\star}= 
\sqrt{{\cal V}} \left(1 \pm 
\sqrt{1 -\left(\frac{v_i \tilde{b}}{v_\star {\cal V}}\right)^2}\right)
\eeq
where ${\cal V}=3/2+v_i^2/(2 v_\star^2)\simeq 3/2$. 
These expressions are obtained by equating the effective potential (including the angular momentum) to zero. The eccentricity $\varepsilon$ is defined as
\beq
\varepsilon=\sqrt{1-\left(\frac{\alpha_-}{\alpha_+}\right)^2}\;.
\eeq

\begin{figure}[h!]
    \centering
    \includegraphics[width=\columnwidth]{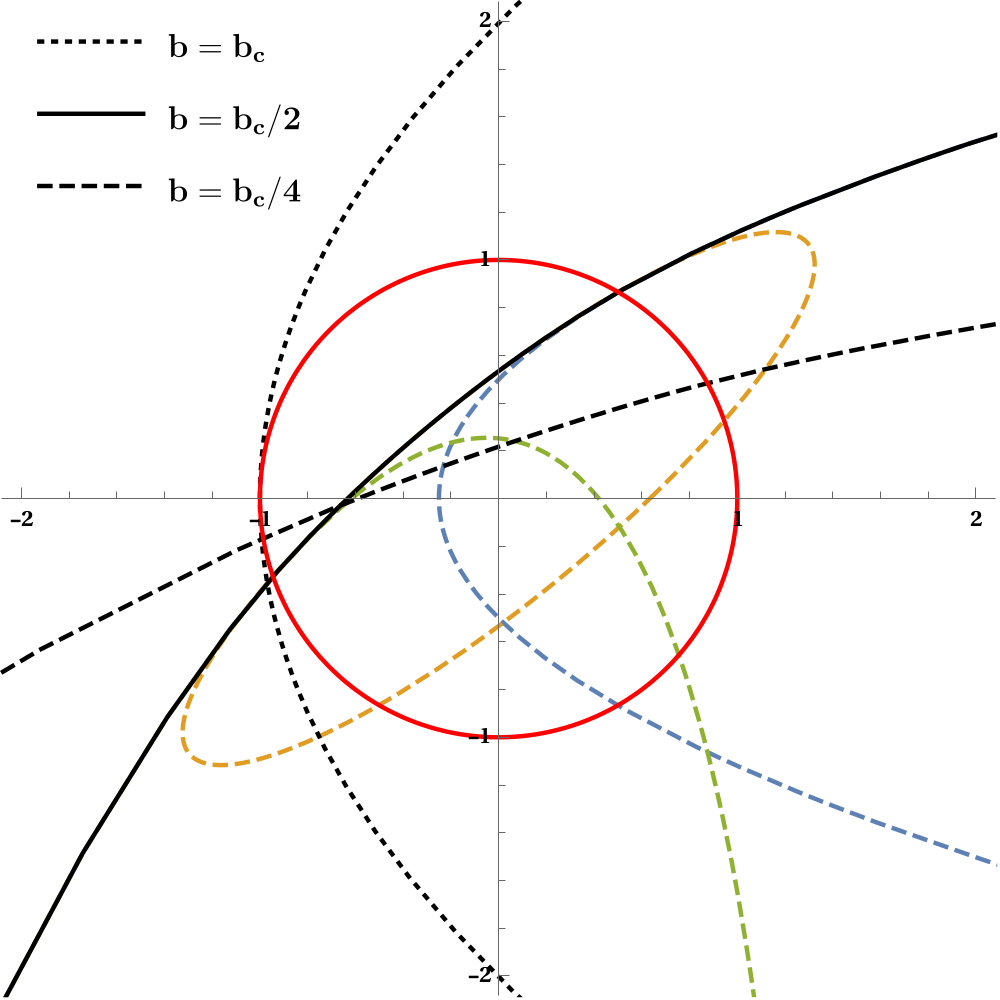}
    \caption{Examples of PBH trajectories (black lines) for impact parameters $\leq b_c$. The star is displayed in red, the construction of this trajectory (black) from two hyperboles (green and blue) and one ellipse (dashed orange) is shown with dashed lines. The radial scale is in units of $R_{\star}$.}\label{fig:motion}
\end{figure}

\begin{figure}[h!]
    \centering
\includegraphics[width=\columnwidth]{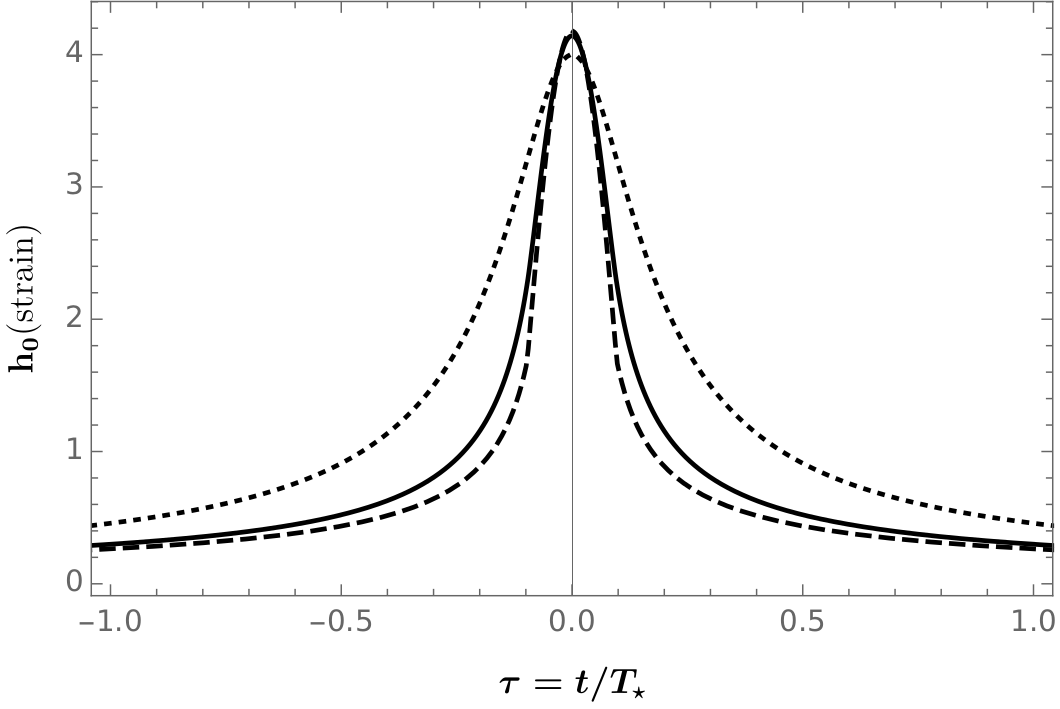}
\caption{Evolution of the gravitational strain for the trajectories shown Fig.~\ref{fig:motion}. $h_c$ is in units of $10^{-25}$, assuming $v_i=10^{-3}$, $d=1\,$kpc and $m=10^{25}\,$g.}\label{fig:gwte}
\end{figure}

A representation of the trajectory can be found in Fig.~\ref{fig:motion} for different impact parameters. For $b=b_c/2$ the two hyperbolas followed by the PBH outside the star (whose border is the red circle) are drawn in blue and green, while the arc of ellipse followed inside the star is shown in dashed orange. Further details on the parameterization of the trajectory with respect to time are given in Appendix~\ref{app:pbh_motion}. The typical GW strain as a function of time for the same trajectories is described by Eq.~(\ref{eq:gwel}) and Appendix~\ref{app:GW} and plotted in Fig.~\ref{fig:gwte}. One can see that the typical gravitational strain and frequency saturate to $h_c = 4\sqrt{5}v_\star^4 R_\star m /(M_\star\,d)$  and $f_c = f_\star$ corresponding to taking the limit $\varepsilon\to 0$ in Eq.~(\ref{eq:gwin}).  

Assuming $N_\star=10^{9}$ NS in the Galaxy, Eq.~(\ref{encontTOT}) yields a total event rate of
\begin{align}
    \Gamma_\star\,{\cal N}_{\star}\simeq 0.38 \;\left( \frac{\rho_\text{BH}}{{\rm GeV\,  cm}^{-3}}\right) \left( \frac{10^{25}\rm g}{m}\right) \left( \frac{10^{-3}}{\bar v}\right){\rm Myr}^{-1}\;,
\end{align}
which, for $m\lesssim 10^{25}\rm g $, is not dissimilar from the estimated GRB rate in the Galaxy. Not surprisingly, this rate of encounters is large for very low $m$: at constant mass density, lighter PBH are more numerous and thus lend to more frequent encounters. On the other hand, the amplitude is proportional to $m$, hence louder encounters require heavier PBH and are correspondingly more rare.

\subsection{GW background from PBH-NS encounters}\label{sec:gwback}

In the previous section, we focused on the single GW emission from a possibly ``loud'' but rare encounter event. However, if PBH constitute a sizable fraction of the DM, for sub-stellar mass PBH there are many PBH traveling near NS at distances below the typical inter-stellar distances, $b_{\rm max}$. It may be therefore interesting to compute the overall GW signal due to these frequent but soft events.  We will start by considering the signal for a single NS, then generalizing the calculation to a population of $N_\star$ NS, spread out in the Galactic disk of radius $R_{G}$. In order to talk of a stochastic background, the frequency of hyperbolic encounters must be larger than the typical frequency of a single encounter merger. This sets a lower distance, $b_{\rm min}$ (of the order 1 AU for $v=10^{-3}$), for the encounter to contribute to the background. 

To set the relevant scales, let us estimate an order of magnitude of the number of encounters contributing to the extremely low GW frequency of $f=10^{-10}$Hz, corresponding to a typical impact parameter of $b=0.1\,$pc.  Thus, considering a PBH density $\rho_{\rm PBH}=1\,$GeV/cm$^3$, there are $N_\star\times\pi\,b^2\, v_i \,\rho_{\rm PBH}/(m\,f)\approx10^4$ events at the same time in the Galaxy, for $N_\star=10^9$, $v_i=10^{-3}$, and $m={10^{25}\rm g}$. This number scales roughly as $1/f^3$, and becomes ${\cal O}(1)$ for frequencies higher than $\sim 10^{-7}\,$Hz, so that computing the GW background above this limit becomes irrelevant.

In the monochromatic approximation~\footnote{More correctly, the emission should be determined via an integral over the trajectory. Given the rather pessimistic conclusions on the detectability of this signal, we deem the monochromatic approximation sufficient.}, the energy released in a single encounter is given by $E_{\rm enc}(b,v_i,d)=P_{\rm gw}/f_c=\kappa \;h_c^2(b,v_i,d)/f_c(b,v_i)$ with $\kappa$ a proportionality constant.
In differential terms in frequency space,
\begin{equation}
\frac{{\rm d}E_{\rm enc}}{{\rm d}f}=\kappa\frac{h_c^2(d,b,v)}{f_c(b,v)}\delta\left(f-f_c(b,v_i)\right)\,.
\end{equation}
The total signal in the limit of incoherent sum can be written as:
\begin{eqnarray}
&& \left\langle \frac{{\rm d}E_{\rm diff}}{{\rm d} f}\right\rangle= \kappa\int_{V_{\rm MW}}{\rm d}V_{\rm MW}\,n_\star\times \nonumber \\
&& \int {\rm d}^3v\int_{b_{\rm min}}^{b_{\rm max}}{\rm d}b\;2\pi b\,v\frac{{\rm d}^3 n_{\rm BH}}{{\rm d}v^3}\;  \frac{h_c^2(d,b,v)}{f_c(b,v)}\delta(f-f_c(b,v))\nonumber 
\end{eqnarray}
where $n_\star$ is the density of stars as a function of the coordinates, and $V_{MW}$ the Milky Way volume considered. Once the integration over $b$ is performed, the delta function fixes the function $b(f,v)$. 
Computing the integral over $v$ then leads to the following value for the effective strain:
\begin{eqnarray}
    \sqrt{\left\langle h_c^2 \right\rangle}&&\simeq 3\times 10^{-20}\left(\frac{10^{-10}\,\rm Hz}{f}\right)^2\times\nonumber\\
    &&  \sqrt{\frac{N_\star}{10^9}\,\frac{m}{10^{25}\rm g}\,\frac{\rho_{\rm PBH}}{\rm GeV\,cm^{-3}}\,\ln\left(\frac{R_G}{20\rm \;kpc}\cdot\frac{\rm pc}{r_p}\right)}
\end{eqnarray}
where $r_p$ is the distance to the closest pulsar. This number is far below the SKA sensitivity~\cite{2015aska.confE..37J} expected to reach $10^{-16}$ for the effective strain measured at around $10^{-8}\,$Hz. Note that this estimate can be extended to the population of ordinary stars in the Galaxy, since the encounters considered here occur at distances larger than $b_{\rm min}\approx 1\,$AU. However, even taking $N_\star$ two orders of magnitude larger is not sufficient to reach the sensitivity of forthcoming low-frequency GW detectors.

\subsection{GW signature of a trapped PBH}\label{trapped}

In the relatively rare cases where the encounter leads to a capture, the PBH motion is also associated to a GW emission. If captured via GW emission in a highly eccentric orbit outside the NS, the GW emission consists of a few  bursts at each periastron passage (the period being a fraction of  Eq.~(\ref{tgwsettl}) of strain and frequency similar to what computed in Sec.~\ref{sec:gw}.

Once orbiting inside the NS, the GW signal is characterized by Eq.~(\ref{eq:conservation}). Interestingly, the expected emission is {\it monochromatic} with frequency $f_\star\sim$kHz and with a {\it constant amplitude} estimated as 
\beq
h_0= \frac{4\sqrt{2}G}{d c^4}m r^2 \omega_\star^2 \approx 2.5\times 10^{-25}  \left( \frac{m}{10^{25}\rm g}\right) \left( \frac{1\; \rm kpc}{d}\right)\;.
\eeq

This GW strain is sustained during the all accretion process, lasting:
\beq
t_{B}= \frac{ c_s^3\, R_\star^3}{3 \,G^2 \,M_\star\, m}\approx 9 \left( \frac{10^{25}\rm g}{m}\right) \rm hours\;.
\eeq
If accounting for rotation (see Sec.~\ref{eq:postcapture}), the GW strain is enhanced or reduced (depending on the sign of $\Omega$) in the last stages of accretion, according to:
\beq
h_0^R(t) = h_0 \,\left( \frac{m(t)}{m} \right)^{\Omega/\omega_\star} = h_0 \, \left( \frac{1}{1-t/t_B} \right)^{\Omega/\omega_\star} \,.
\eeq

Assuming $N_\star=10^{9}$ neutron stars in the Galaxy, Eq.~(\ref{eq:cap_rate}) yields an event rate of
\beq
{\cal G}_{\star}N_\star\simeq 2.1\times 10^{-8} \; \left( \frac{\rho_\text{PBH}}{ \rm GeV\,cm^{-3}}\right) \left( \frac{10^{-3}}{\bar v}\right)^3 {\cal C}\left[X \right] \rm yr^{-1}\,.\label{eq:cap_rate_MW}
\eeq
For typical Milky Way values of ($\rho_{\rm PBH},m,\bar{v}$), within the age of the Galaxy ($\approx 10^{10}\,$\rm yr) one would expect up to a few hundreds cases of NS transmuted in BH.
Note, however, that provided that $X$ is in the range where ${\cal C}\left[X \right] $ is constant, the capture rate is maximized in environments with large $\rho_{\rm PBH}$ and low velocity dispersion, singling out DM dominated dwarf spheroidals as comparatively more promising targets. Typical such objects (see for instance~\cite{Read:2018fxs}) have a velocity dispersion one order of magnitude or more below the Milky Way value and DM densities one order of magnitude higher than in the solar neighborhood, hence we expect that ${\cal G}_{\star}$ can be enhanced by 10$^4$ or more compared to the Milky Way value. Since each of these objects contains $\sim 10^{-4}$ of the stars of the Milky Way, the overall numbers of NS transmuted in BH may be thus comparable.

\subsection{Final stages}
If a PBH is trapped inside the NS, eventually it will swallow the entire NS, causing a so-called {\it transmutation} of the NS into a BH. This phenomenon is expected to be associated with both electromagnetic (EM) and gravitational wave signals. The reason why some EM burst is expected boils down to the 
no-hair theorem and the fact that NS are magnetized objects: The newly formed BH must expel its magnetic field energy, liberating at least an energy~\cite{Chirenti:2019sxw}
\begin{equation}
E_B=\frac{B^2}{8\pi}\frac{4\pi}{3}R_\star^3 \simeq 2\times 10^{41}\left(\frac{B}{10^{12} {\rm G}}\right)^2 \left(\frac{R_\star}{10\, {\rm km}}\right)^3{\rm erg}\;
\end{equation}
into EM form. For some more details, see \cite{2015MNRAS.450L..71F,2018ApJ...868...17A,Chirenti:2019sxw}. It is unclear if further signatures are associated to the ejecta, if present in non-negligible amounts.  Also, a fast change of the quadrupole will lead to some GW signature. These signals have been estimated to be rather unpromising for detection~\cite{2019arXiv190907968E} (see also~\cite{2011PhRvD..83h3512K,2019JCAP...05..035G}). However,  current simulations have set the PBH exactly at the center, forcing a symmetry which definitely suppresses both the GW emission and other signatures (e.g. ejecta), and realistic magnetic fields are not accounted for. Our study (and notably  Eq.~(\ref{initcond})) suggests some degree of asymmetry in the final phase of the PBH mass growth, which is more and more pronounced for a heavier and heavier PBH. For instance, a PBH of initial mass $\simeq 10^{-3}\,M_\odot$ will have reached a mass of 10\% of the NS (which one may consider  at the onset of the final transmutation) at a distance of about 10\% of the NS center (about 3-4 times larger than its Schwarzschild radius). Although we cannot compute reliably the signatures associated to the final stages, a relation like Eq.~(\ref{initcond}) can be used to provide a more realistic initial condition in future simulations.


\section{Conclusion} \label{concl}
In this article, we have revisited the interaction processes between  primordial black holes (PBH) and neutron stars (NS) and discussed their consequences for the dynamical evolution of the system. 

In particular, we have
argued that dynamical friction, the major player in the PBH capture (which happens typically with the PBH hitting the NS at supersonic velocities), is negligible in the post-capture dynamics, when the PBH is orbiting within the star at subsonic speed.

Also, we have shown that (Bondi-like) accretion dominates the post-capture phase, and is responsible for an approximate conservation law, valid until the final stages, when the transmutation of the NS into a BH takes place. This also implies that the  onset of the final catastrophic event is expected with the BH seed in slight off-center position: While the actual consequences of this fact must be investigated via numerical simulations, one can expect enhanced  electromagnetic and gravitational wave signatures compared to current estimates. 

For the first time, we also assessed the importance of GW losses in this context, notably for captures at large impact parameters in low velocity dispersion systems. 

Finally, we discussed GW signals associated to different phases of the PBH-stellar interaction. In particular, we extended the hyperbolic encounter ``tear drop'' signal calculation to the case where the PBH enters the NS in its trajectory, and estimated the (small) GW background from frequent soft encounters.  Unfortunately, for the single encouter case the signal rate and strength are anticorrelated: We expect sufficiently loud events (associated to massive PBH) to be rare, while frequent events (for light PBH) are below current or foreseen GW sensitivity. 
Barring some luck, the still uncertain emission associated to the transmutation event appears the most promising opportunity for a discovery of these exotics. 

It is interesting however to point out that as the result of cumulative transmutation events over the cosmic history, a population of low-mass BH (with mass $\sim 1\div \, M_\odot$) will build up. It has been speculated that up to a few percent of the NS-NS coalescence events may in fact involve such a transmuted low-mass BH~\cite{2018ApJ...868...17A}. This promising alternative diagnostics will however require high GW event statistics and a good measurement of the merger/ringdown part of the waveform, for which one will have to wait for third-generation GW detectors~\cite{Yang:2017gfb}.


\begin{acknowledgments}
  Y.G warmly thanks Nicolas Chamel for providing him the equations of state for old neutron stars and for discussions. The work of Y.G. is supported by  Villum Fonden under project no.~18994. PDS acknowledges support from IDEX Univ. Grenoble Alpes, under the program {\it Initiatives  de  Recherche Strat{\'e}giques}, project ``Multimessenger avenues in gravitational waves'' (PI: PDS). The work of P.T. is supported in part by the IISN grant 4.4503.15. 
\end{acknowledgments}


\begin{appendix}
\section{Details on PBH trajectory}\label{app:pbh_motion}
\subsection{Parameterization of the PBH motion}
In this appendix we give the parameterization of the PBH trajectory for impact parameter $b<b_c$ (see Eq.~\ref{eq:bc}) for which the PBH crosses the NS. The case $b>b_c$ can readily be deduced for example from Refs.~\cite{2018PDU....21...61G,2017PDU....18..123G}. In the following we consider the classical trajectory of a PBH of mass $m\ll M_\star$ crossing a NS of constant density $\rho_\star=3M_\star/(4\pi R_\star^3)$. Using the polar coordinates ($r,\phi$), with $r=0$ corresponding to the NS center, the trajectory is parameterized by a hyperbola (I), an ellipse (II) and a hyperbola (III):
\beq
r(\phi)=
\begin{cases}
r_{\rm \tiny I}(\phi)=\displaystyle\frac{a (e^2-1)}{1+e\cos(\phi-\psi_0)}, &\phi\leqslant\phi_0\\[10pt]
r_{\rm \tiny II}(\phi)=\displaystyle\frac{\alpha_-}{\sqrt{1-\varepsilon^2\cos^2(\phi-\psi_1)}}, &\phi_0<\phi<\phi_1\\[10pt]
r_{\rm \tiny III}(\phi)=\displaystyle\frac{a (e^2-1)}{1+e\cos(\phi-\psi_2)}, &\phi_1\leqslant\phi\;,\\
\end{cases}
\eeq

For the hyperbolic motion ($r_{\rm I}$ and $r_{\rm III}$), the eccentricity is given by:
\beq
e=\sqrt{1+\frac{b^2}{a^2}}=\sqrt{1+\tilde{b}^2\left(\frac{v_i}{v_\star}\right)^4}\;.
\eeq
with semi-major axis $a_h$ such that:
\beq
\left(\frac{a}{R_\star}\right)^2 = \left(\frac{v_\star}{v_i}\right)^4\;.
\eeq 
For the ellipsoid motion ($r_{\rm II }$), as recall in the main text, the eccentricity is given by:
\beq
\varepsilon=\sqrt{1-\left(\frac{\alpha_-}{\alpha_+}\right)^2}\;,
\eeq
with the semi-major and semi-minor axis $\alpha_\pm$ such that,
\beq
\tilde{\alpha}_\pm\equiv \frac{\alpha_\pm}{R_\star}= 
\sqrt{{\cal V}} \left(1 \pm 
\sqrt{1 -\left(\frac{v_i \tilde{b}}{v_\star {\cal V}}\right)^2}\right)\,,
\eeq
where ${\cal V}=3/2+v_i^2/(2 v_\star^2)\simeq 3/2$. 

Concerning the angles of the problem, while $\psi_0$ is commonly defined as:
\beq
\psi_0 = \arccos[-1/e]\;,
\eeq
the other angles are obtained requiring the continuity of the trajectory:
\begin{align}
[r_{\rm I}(\phi_0)=R] \Rightarrow \phi_0 &= \psi_0 - \arccos\left[\frac{1}{e} \left(\tilde{b}\sqrt{e^2 - 1} - 1\right)\right]
\\ 
[r_{\rm II}(\phi_0)=R] \Rightarrow \psi_1 &= \phi_0 - \arccos\left[\frac{1}{\varepsilon} \sqrt{1 - \tilde{\alpha}_-^2}\right]\;,
\end{align}
and by symmetries,
\begin{align}
\phi_1 &= \pi - \phi_0 + 2\psi_1 \, \\
\psi_2 &= 2\psi_1 - \psi_0 + \pi\;.
\end{align}

The time evolution $\tau=t(\phi)/T_\star$ can also be split in the same $\phi$ intervals such that,
\beq
\tau(\phi)=
    \begin{cases}
    \tau_{\rm \tiny I}(\phi), &\phi\leqslant\phi_0\\[10pt]
    \tau_{\rm \tiny II}(\phi), &\phi_0<\phi<\phi_1\\[10pt]
    \tau_{\rm \tiny III}(\phi), &\phi_1\leqslant\phi\;,\\
    \end{cases}
\eeq
with the recursive definitions,
\begin{align}
\tau_{\rm \tiny I}(\phi) &= {\cal T}_{\rm \tiny out}(\phi - \psi_0)-{\cal T}_{\rm \tiny out}(\phi_0 - \psi_0)\\
\tau_{\rm \tiny II}(\phi) &= {\cal T}_{\rm \tiny in}(\phi-\psi_1)-{\cal T}_{\rm \tiny in}(\phi_0-\psi_1)+\tau_{\rm \tiny I}(\phi_0)\\
\tau_{\rm \tiny III}(\phi) &= {\cal T}_{\rm \tiny out}(\phi - \psi_2)-{\cal T}_{\rm \tiny out}(\phi_1 - \psi_2) + \tau_{\rm \tiny II}(\phi_1)\;,
\end{align}
where we have introduced the functions:
\begin{align}
{\cal T}_{\rm \tiny out}(u)= \frac{\tilde{b}}{2 \pi}\frac{v_\star}{v_i}\, &\Bigg(\;\frac{e\,\sin u}{1 + e\,\cos u} \\ &- \frac{2}{\sqrt{e^2 -1}} \tanh^{-1}\left[\sqrt{\frac{e - 1}{e + 1}} \tan{\frac{u}{2}}\right]\Bigg)\;, 
\end{align}
and,
\beq
{\cal T}_{\rm \tiny in}(u)= \; \frac{\alpha_-^2}{2\pi\,b\,R_\star}\frac{v_\star}{v_i}\frac{1}{\sqrt{1-\varepsilon^2}} \tan^{-1}\left[\frac{1}{\sqrt{1-\varepsilon^2}} \tan{u}\right]\;.\\ 
\eeq

\subsection{Gravitational wave emission}\label{app:GW}
The function used to compute the GW  strain $h_0$ Eq.~(\ref{eq:gw_general}) is a piecewise function depending on the regime of the motion:    
\beq
g(\phi)=
    \begin{cases}
    g_{\rm \tiny out}(\phi-\psi_0), &\phi\leqslant\phi_0\\[10pt]
    g_{\rm \tiny in}(\phi-\psi_1), &\phi_0<\phi<\phi_1\\[10pt]
    g_{\rm \tiny out}(\phi-\psi_2), &\phi_1\leqslant\phi\;,\\
    \end{cases}
\eeq
with,
\beq
\begin{aligned}
g_{\rm \tiny out}(\phi)&=  \frac{\sqrt{2}}{3}\frac{1}{e^2-1}\left[36+59 \,e^2+10 \,e^4  \right.\\ & \left. +(108 + 47 \,e^2)\,e\cos{\phi}+59\,e^2\cos{2\phi}+9\,e^3\cos{3\phi}\right]^{1/2}
\end{aligned}
\eeq
and,
\beq
\begin{aligned}
g_{\rm \tiny in}(\phi)= & \frac{{2}}{3}\left(\frac{b}{\alpha_-}\right)^2 \big[ 38\,(1-\varepsilon^2) +5\,\varepsilon^4   \\ &  +80 (1-\varepsilon^2)^2 \left(2-\varepsilon^2(1+\cos{ 2\phi} \right)^{-2} \\  & +40 (2-3\,\varepsilon^2 + \varepsilon^4)\left(2-\varepsilon^2(1+\cos{2\phi})\right)^{-1} \big]^{1/2}\;. \label{eq:gwin}
\end{aligned}
\eeq

The power radiated in gravitational waves can be computed from \cite{1918SPAW.......154E}, as:
\beq
P_{\rm gw}=\frac{d E_{\rm gw}}{dt}=-\frac{G}{5 c^5}\langle \dddot Q_{ij} \dddot Q^{ij} \rangle
\eeq
We define the differential energy radiated by unit angle $p^\phi_{\rm gw}$ as:
\beq
p^\phi_{\rm gw}=\;\frac{d E_{\rm gw}}{d\phi} \;=\frac{1}{\dot \phi}\;P_{\rm gw}\;. 
\eeq
We compute this quantity for the PBH travelling in or out of the star:

\beq
p^\phi_{\rm gw}(\phi,\tilde b)= \frac{2}{45} E_i \frac{v_\star^2 v_i^3}{c^5 } \frac{m}{M_\star}
    \begin{cases}
    f_{\rm \tiny out}(\phi-\psi_0), &\phi\leqslant\phi_0\\[10pt]
    f_{\rm \tiny in}(\phi-\psi_1), &\phi_0<\phi<\phi_1\\[10pt]
    f_{\rm \tiny out}(\phi-\psi_2), &\phi_1\leqslant\phi\;,\\
    \end{cases}
\eeq
with,
\begin{align}
f_{\rm out} (\phi, \tilde b)= \frac{2}{\tilde b}\frac{(1 + e \cos\phi )^2}{(e^2 -1)^3} &(\,144 + 288 \,e\, \cos\phi \nonumber \\ & + 77 \,e^2  + 67\, e^2 \cos 2 \phi\, ) \;, 
\end{align}
and,
\begin{align}
f_{\rm in} (\phi, \tilde b)= 
4 \frac{\tilde b^5}{\tilde{\alpha}_-^6}& \frac{(1 - \varepsilon^2)^2}{(1  - \varepsilon^2\cos^2 \phi)^3}  (\,72\,(1- \varepsilon^2) + 37 \,\varepsilon^4 \nonumber \\ & + 36\, \varepsilon^2\, (\varepsilon^2-2 ) \,\cos 2\phi  - \varepsilon^4 \cos 4 \phi \,)\;.
\end{align}
From these expressions on can compute the gravitational energy radiated outside the NS,  
\beq
|\Delta E|_{\rm gw}^{\rm out} (\tilde b) = \int^{\phi_0}_0 p^\phi_{\rm gw}(\phi,\tilde b) \, d\phi +\int_{\phi_1}^{\phi_{dev}} p^\phi_{\rm gw}(\phi,\tilde b) \, d\phi\;,
\eeq
with $\phi_{dev}=2\psi_1$ the total deflection angle, and inside the NS,
\beq
|\Delta E|_{\rm gw}^{\rm in} (\tilde b)  =\int^{\phi_1}_{\phi_0} p^\phi_{\rm gw}(\phi,\tilde b) \, d\phi\;.
\eeq

\end{appendix}

\bibliographystyle{apsrev4-1}
\bibliography{pbh_capture}
\end{document}